\documentclass[manuscript]{acmart}

\AtBeginDocument{%
  \providecommand\BibTeX{{%
    \normalfont B\kern-0.5em{\scshape i\kern-0.25em b}\kern-0.8em\TeX}}}

\copyrightyear{2021} 
\acmYear{2021} 
\setcopyright{rightsretained} 
\acmConference[RecSys '21]{Fifteenth ACM Conference on Recommender Systems}{September 27-October 1, 2021}{Amsterdam, Netherlands}
\acmBooktitle{Fifteenth ACM Conference on Recommender Systems (RecSys '21), September 27-October 1, 2021, Amsterdam, Netherlands}
\acmDOI{10.1145/3460231.3474616}
\acmISBN{978-1-4503-8458-2/21/09}

\begin{document}

\title{Scaling Enterprise Recommender Systems for Decentralization}

\author{Maurits van der Goes}
\email{maurits.vandergoes@heineken.com}
\orcid{0003-3743-5588}
\affiliation{%
  \institution{The HEINEKEN Company}
  \city{Amsterdam}
  \country{Netherlands}
}

\renewcommand{\shortauthors}{van der Goes}

\begin{abstract}
  Within decentralized organizations, the local demand for recommender systems to support business processes grows. The diversity in data sources and infrastructure challenges central engineering teams. Achieving a high delivery velocity without technical debt requires a scalable approach in the development and operations of recommender systems. At the HEINEKEN Company, we execute a machine learning operations method with five best practices: pipeline automation, data availability, exchangeable artifacts, observability, and policy-based security. Creating a culture of self-service, automation, and collaboration to scale recommender systems for decentralization. We demonstrate a practical use case of a self-service ML workspace deployment and a recommender system, that scale faster to subsidiaries and with less technical debt. This enables HEINEKEN to globally support applications that generate insights with local business impact.
\end{abstract}




\maketitle

\section{Introduction}
The interest in data analytics applications in enterprises grows as studies show it leads to an increase in productivity \cite{muller2018effect}. Within enterprises, numerous large datasets are available \cite{foorthuis2020algorithmic}. However, their current software is unable to process them for new insights. Like a recommender system that advises sales employees for their outlet visit schedule by modeling historic data. The HEINEKEN Company is the most global brewer with breweries in more than 70 countries. With global support, these local subsidiaries build and exchange successful recommender systems. These systems with domestic impact quickly gather interest from other subsidiary brewers. However, its international and decentralized structure also comes with diversity in data sources and information technology. Meeting the expected demand from subsidiaries for recommender systems requires a scalable approach in development and operations that decentralizes the ownership of the system.

Growing the data engineering team is expensive and demanding with the shortage of data talent \cite{zhang2021ai}. Furthermore, studies show that increasing team size does not linearly increase the software efforts and quality \cite{pieterse2006software,pendharkar2007empirical}. To enable scaling is vital to achieve efficiency in the design, development, and operations of recommender systems. Immature infrastructure in production causes long-term costs expressed as technical debt \cite{cunningham1992wycash,sculley2015hidden}. With the risk of bringing  organizations to a stand-still. To stimulate delivery velocity and prevent technical debt while scaling the recommender system, this efficiency requires to be included from the initial design. This talk discusses a software engineering method for scalable recommender systems, two applications, and its learnings.

\section{Approach}
\begin{figure*}[ht]
  \centering
  \includegraphics[scale=0.3]{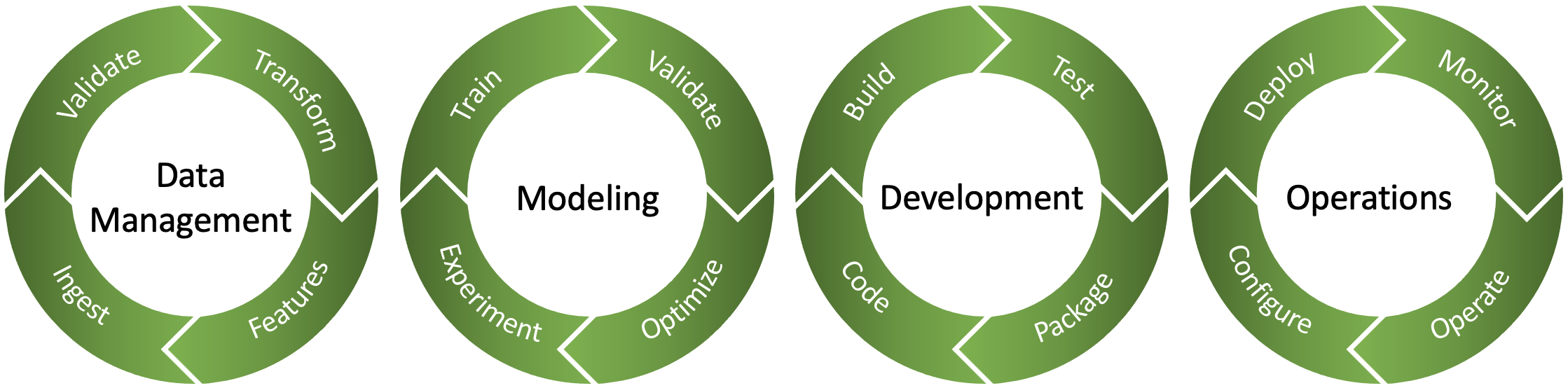}
  \caption{Machine Learning Operations life cycle.}
  \Description{Machine Learning Operations life cycle in four phases: data management, modeling, development and operations.}
  \label{figure:mlops-lifecycle}
\end{figure*}

Compared to software engineering technical debt is a larger challenge for recommender systems as code is accompanied by datasets and machine learning models. We draw from experience in addressing this challenge with a Machine Learning Operations (MLOps) method. MLOps is the combination of machine learning and development operations (DevOps). The goal of the MLOps practices is to bring machine learning applications into production with attention to reproducibility, reliability, and efficiency \cite{lwakatare2020devops}. There are two main variants of this MLOps life cycle with either three (excluding data management) or four stages. We choose the four-stage life cycle (fig.~\ref{figure:mlops-lifecycle}) since proper data management is essential for successful recommender systems \cite{hulsebos2019sherlock}.

The MLOps method is executed five best practices:
\begin{itemize}
    \item \textbf{Pipeline automation:} Removing manual actions and delays by automating actions and connecting services in pipelines. For example by releasing solutions with a continuous integration and continuous deployment pipeline. 
    \item \textbf{Data availability:} Access to validated datasets via a feature store and indexed in a data catalog to ensure reproducible machine learning.
    \item \textbf{Exchangeable artifacts:} All machine learning models, code, and configurations are version controlled, descriptive, and PEP8 consistent. Solution patterns with documentation are available for common architecture artifacts, like integrating external systems. Code scripts are preferred over notebooks in production. Notebooks have upsides for experimenting, but these do not exceed the advantages of regular scripts. 
    \item \textbf{Observability:} Deep understanding of the system components to ensure performance and identify root causes of issues. Metrics, events, logs, and traces are collected and accessible to the product team, not only engineers.
    \item \textbf{Policy-based security:} Releases are separated into four environments: development, testing, acceptance, and production. Authorization to these environments is managed with Attribute Based Access Control (ABAC) with control plane and data plane functions. Secrets are stored in environment-specific key vaults.
\end{itemize}
This MLOps method and practices are a shared responsibility of both data scientists and data engineers. Committing to the same method enables the \textit{“You build it, you run it”}-principle by Werner Vogels \cite{o2006conversation}. Advantageous for delivery velocity and quality of the applications.

\section{Applications}
We selected two cloud native applications on the Azure platform to present: a self-service ML workspace deployment and a recommender system. These are intertwined, as the ML workspace is the engine of the recommender system. Together these applications are designed globally to support local analytics demand in a scalable form with distributed ownership.

\subsection{Self-service ML workspace deployment}
\begin{figure}[ht]
  \centering
  \includegraphics[scale=0.4]{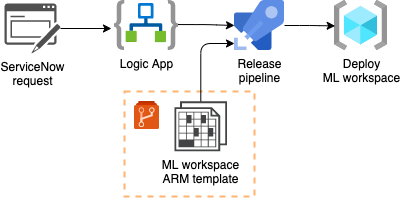}
  \caption{The architecture of the pipeline to automatically deploy an Azure ML workspace.}
  \Description{The architecture consisting of a ServiceNow and Azure resources.}
  \label{fig:arch-workspace}
\end{figure}

An Azure ML workspace is a cloud workbench for machine learning development and deployment. It empowers the data scientist with a self-service environment that contains various MLOps features and abstracts away technical complexities like containerization. Requesting and delivering a new ML workspace was a manual and time-consuming process. With straightforward technology, this is effectively automated (fig.~\ref{fig:arch-workspace}). The request submitted via ServiceNow is sent to the HTTP endpoint of the Azure Logic App. ServiceNow is familiar software for the end-users. In Logic App the configuration details are parsed and sent via a REST API to an Azure DevOps Release pipeline. Using an Azure Resource Manager (ARM) template with the configuration the ML workspace is deployed in the designated Azure environment within minutes.

\subsection{Recommender system}
\begin{figure*}[ht]
  \centering
  \includegraphics[scale=0.4]{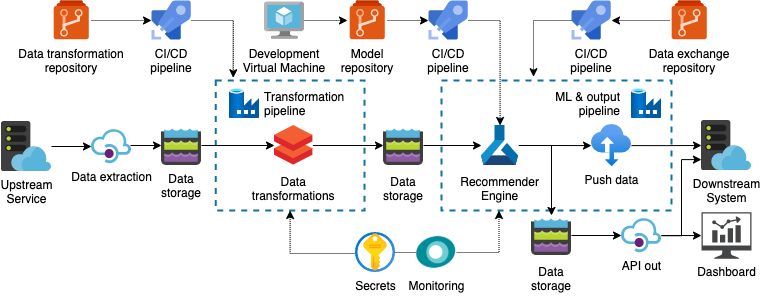}
  \caption{The architecture of a recommender system with Azure resources.}
  \Description{The architecture consisting of various Azure Cloud and DevOps resources.}
  \label{fig:arch-recsys}
\end{figure*}

The recommender system has a decoupled architecture, consisting of various solution patterns (fig.~\ref{fig:arch-recsys}). Data is extracted from upstream services (e.g. SAP or Salesforce) and stored in the Azure data lake. The ML workspace provides functionality for both model research, running pipelines as recommender engine, and tracking performance. The Azure Data Factory pipeline first triggers the ML pipeline and subsequently pushes the data to a downstream service (dashboard, external system API, or internal API). All the steps of this recommendation process are scheduled or triggered on data events. With localization in mind, the configurations are stored in separate yaml files. All resources templates and configurations are stored and deployed as infrastructure as code from individual Azure DevOps repositories.

\section{Learnings}
Analyzing the practical case with the implementation of MLOps generates learnings. The engineering bottlenecks for scaling recommender systems are diversity in data sources and information technology with subsidiaries. This challenge is successfully diminished by engineering a culture that fosters self-service, automation, and collaboration. The ML workspace deployment application is an effective example of self-service and automation. This deployment capability is expanded from a few engineers to all data colleagues by the implementation of a pipeline of connected services. As a result, valuable engineering time is freed up and the delivery time of an ML workspace to its end-users is reduced from weeks to minutes. Conveniently onboarding data scientists to the ML workspace with central knowledge resources further contributes to a higher velocity.

The presented recommender system is designed with internal solution patterns, automated deployment of resources, centrally available datasets, and accessible observability. This decreases the engineering time per subsidiary use case while keeping the agility to effectively incorporate local specifications. By dealing with the diversity in data sources and information technology early and with solution patterns, it is possible to reuse many recommender system artifacts across subsidiaries. Substantially reducing the collective technical debt of the organization. From our experience, resulting in scaling recommender systems faster to multiple countries. The success of a local recommender system is determined by multiple KPIs, including revenue, savings, usage, and user valuation. The attention for standardization, exchangeable artifacts, and available datasets leads to more collaboration between colleagues across teams or country borders. By promoting a culture of self-service, automation, and collaboration, the HEINEKEN Company is better qualified to support its subsidiaries with recommender systems or other analytics requests. 

\section{Conclusion}
Meeting the local demand of recommender systems in enterprises requires efficient engineering with minimal technical debt. This talk described an MLOps method and two corresponding applications. We learned that a culture with self-service, automation, and collaboration is key to enable an enterprise to scale recommender systems with shared ownership. We believe that these engineering principles are valuable to other enterprises with similar data-driven ambitions and challenges. Additionally, we think that these are not bounded to recommender systems, but applicable for scaling various types of machine learning applications globally.

\begin{acks}
This talk stands on the shoulders of talented people (at The HEINEKEN Company). Their ideas, code, and feedback shaped our thoughts and contributed to this work.
\end{acks}

\section{Speaker Bio}
\textbf{Maurits van der Goes} is currently a Data Engineer in the Global Analytics team at The HEINEKEN Company, where he specializes in their MLOps infrastructure for e.g. information filtering pipelines in sales, marketing, logistics, and finance. Earlier at RTL Netherlands, he developed the first news recommender system for major Dutch website RTL Nieuws. Maurits graduated from Delft University of Technology with a Masters in Complex IT Systems Engineering and Management. His graduation research was on a team-formation recommender system that he developed at a digital platform for self-managing teams.

\bibliographystyle{ACM-Reference-Format}
\bibliography{references}

\end{document}